\documentclass[runningheads]{llncs}

\usepackage{graphicx}
\usepackage{amsmath}
\usepackage{algorithm}
\usepackage{algpseudocode}
\usepackage{subfig}
\usepackage{tabularx}
\usepackage{placeins}
\usepackage{enumitem}

\usepackage{array}
\newcolumntype{P}[1]{>{\centering\arraybackslash}p{#1}}

\usepackage{caption}
\usepackage[a4paper, margin=1.1in]{geometry}

\newcolumntype{C}{>{\centering\arraybackslash}X}

\begin{document}

\title{Relay Mining: Incentivizing Full Non-Validating Nodes Servicing All RPC Types}
\titlerunning{Relay Mining}

\author{Olshansky, Daniel\inst{1}
\and  Rodr\'iguez Colmeiro, Ramiro\inst{2}
}

\authorrunning{D. Olshansky, R. Rodr\'iguez Colmeiro}
\institute{
    Grove Inc.;  \url{grove.city}
    \and
    Pocket Scan Technologies LLC; \url{poktscan.com}
}

\maketitle 

\begin{abstract}

Relay Mining presents a scalable solution employing probabilistic mechanisms, crypto-economic incentives, and new cryptographic primitives to estimate and prove the volume of Remote Procedure Calls (RPCs) made from a client to a server. Distributed ledgers are designed to secure permissionless state transitions (writes), highlighting a gap for incentivizing full non-validating nodes to service non-transactional (read) RPCs. This leads applications to have a dependency on altruistic or centralized off-chain Node RPC Providers. We present a solution that enables multiple RPC providers to service requests from independent applications on a permissionless network. We leverage digital signatures, commit-and-reveal schemes, and Sparse Merkle Sum Tries (SMSTs) to prove the amount of work done. This is enabled through the introduction of a novel ClosestMerkleProof proof-of-inclusion scheme. A native cryptocurrency on a distributed ledger is used to rate limit applications and disincentivize over-usage. Building upon established research in token bucket algorithms and distributed rate-limiting penalty models, our approach harnesses a feedback loop control mechanism to adjust the difficulty of mining relay rewards, dynamically scaling with network usage growth. By leveraging crypto-economic incentives, we reduce coordination overhead costs and introduce a mechanism for providing RPC services that are both geopolitically and geographically distributed. We use common formulations from rate limiting research to demonstrate how this solution in the Web3 ecosystem translates to distributed verifiable multi-tenant rate limiting in Web2.


\end{abstract}

\section{Introduction}

The rise of Software as a Service (SaaS) has popularized subscription-based models, facilitated through Remote Procedure Calls (RPCs) like JSON-RPC, gRPC, REST APIs, and GraphQL to meet diverse application requirements. These models typically bill development teams based on usage or subscription tiers \cite{salesforceWhatSaaS}. 

In the Web3 space, decentralized applications (dApps) often rely on blockchain data, traditionally accessed by operating a full node \cite{alchemyProsCons}. However, the high resource demands — such as the need for at least 16GB of RAM, a 2TB SSD, and a 25Mbps download speed for running a geth client \cite{ethereumHardwareRequirements} — make self-managing full nodes impractical for most developers, particularly on client devices. Light clients, a concept originating from Bitcoin's Simple Payment Verification, are evolving but still depend on Node RPC providers that supply necessary proofs and headers \cite{nakamoto2008bitcoin}.

The dependence on these providers accelerates dApp development but also pushes towards further centralization and reliance on a few large infrastructure entities \cite{moxie2022}. While light clients aim to minimize this reliance, they still require RPC services to function effectively.

For sustainability, RPC service operators implement rate-limiting and Denial-of-Service (DoS) protection strategies. Despite advancements in distributed ledger technology for permissionless transaction verification (\textbf{writes}), a corresponding solution for data reads (\textbf{reads}) remains undeveloped. Relay Mining addresses this by exploring rate-limiting mechanisms within a live network, applicable to any RPC interface.

\subsection{Blockchain Writes - Validators, Builders, Proposers et al}

Distributed ledgers (blockchains) are often reduced to the problem of Byzantine Fault Tolerant (BFT) State Machine Replication (SMR) \cite{buchman2019latest}. Discussions about scalability typically focus on increasing throughput — the capacity to write more data to the ledger without compromising \textbf{safety} and \textbf{liveness}. This challenge led to the formulation of the Blockchain Scalability Trilemma, which assesses the trade-offs between scalability, decentralization, and security of a blockchain \cite{Halpin_2020}.

When evaluating the scalability of a Layer 1 (L1) blockchain, key metrics include Transactions-Per-Second (TPS) and Time-to-Finality (TTF), which provide direct insights into the network's capacity and usage, as illustrated in Table ~\ref{l1-throughput}.

The constraints of limited write capacity lead to on-chain auctions, often analyzed through the lens of Maximal Extractable Value (MEV). The need for efficient transaction ordering has introduced additional roles like Proposers and Builders, enhancing the ecosystem's focus on scaling data throughput rather than data consumption \cite{kulkarni2023theory} \cite{heimbach2023ethereums}.

\begin{table}
    \centering    
    \begin{tabular}{|l|P{1.5cm}|P{1.2cm}|P{1.2cm}|P{1.1cm}|P{1cm}|}
        \hline & Ethereum & Solana & Dfinity & Aptos & Near \\
        \hline
            Transactions-Per-Second (TPS) & 11 & 286 & 5382 & 5 & 8.25  \\
            Time-To-Finality (TTF) &  15m & 5-12s & 1.4s & $\le$ 1s & 3.3s \\
            Number of Nodes &  6562 & 1872 &  823 & 107 & 798 \\
        \hline
    \end{tabular}
    \caption{L1 write throughput comparison measured in December 2022; Aptos data is more recent \cite{internetcomputerComparisonInternet} \cite{aptoslabsAptosExplorer}.}
    \label{l1-throughput}
\end{table}

\subsection{Blockchain Reads - RPC Node Providers}

Despite advancements in addressing the Scalability Trilemma to manage, verify, and incentivize writes, the predominant type of requests made by application developers are \textbf{reads}. Table.~\ref{rpc-node-usage} highlights the substantial traffic managed by major node providers in 2023, emphasizing the billions of daily requests, mostly reads, which present challenges not addressed by BFT replicated state machines.

Accessing blockchain data, crucial for operations like those described in the Simple Payment Verification \cite{nakamoto2008bitcoin}, typically requires one of the following infrastructure components:
\begin{enumerate}
    \item \textbf{Full Nodes}: Synchronizing a full node to maintain a complete, locally verified copy of the ledger.
    \item \textbf{Light Clients}: Acquiring headers and proofs just-in-time from other full nodes.
    \item \textbf{RPC Providers}: Relying on a third party to retrieve data via an RPC request.
\end{enumerate}

\begin{figure}
    \centering
    \includegraphics[width=0.7\textwidth]{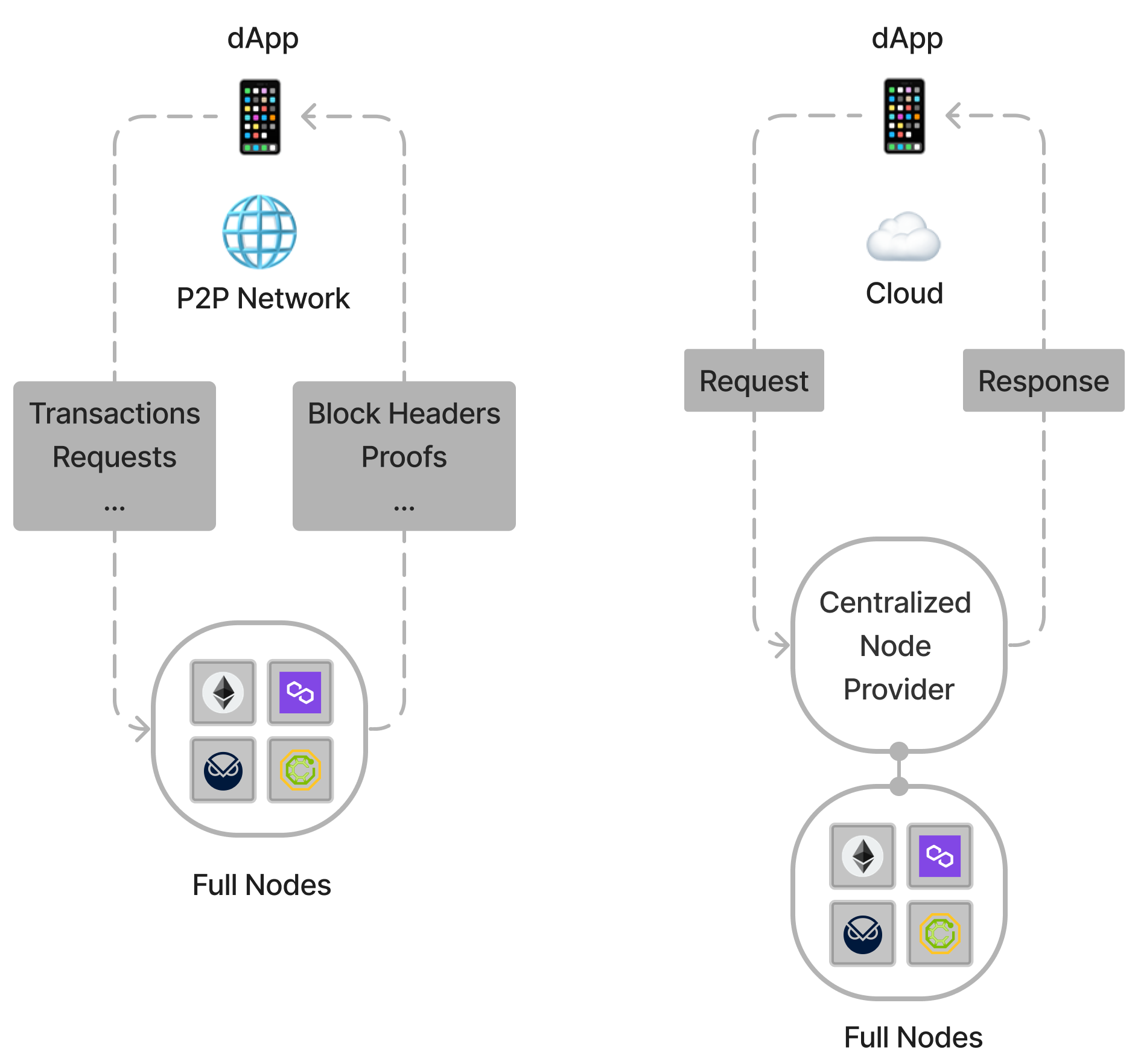}
    \caption{Communication between a dApp with permissionless and untrusted full nodes (left) versus trusted and centralized full nodes (right).} \label{fig:light-full-node}
\end{figure}

Options (1) and (2) often face practical limitations for widespread application use. Although option (3) has evolved into a robust Web2 SaaS-like ecosystem, it depends on provisioning and maintaining scalable infrastructure, while managing off-chain payments and Service Level Agreements (SLAs) in a more traditional and permissioned environment. Fig.~\ref{fig:light-full-node} illustrates the operational models commonly adopted by decentralized Applications (dApps) today using these methods.

\begin{table}
    \centering    
    \begin{tabular}{|l|P{1.5cm}|P{1.2cm}|P{1.2cm}|P{1.1cm}|P{2cm}|}
        \hline & Infura & Alchemy & Ankr & Pocket Network \\
        \hline
            Requests Per Day & 7B & ?? & 8B & 1.5B  \\
            Latency / Round-Trip-Time (sec) & 0.77 & 0.61 & 0.67 & 0.76  \\
            Availability / Uptime & 99.99\% & ?? & 99.99\% & 99\%  \\
        \hline
    \end{tabular}
    \caption{The latency shown is the p50 RTT for Ethereum Mainnet from Miami as measured using RPCMeter ~\cite{rpcmeterRPCMeter}. The other metrics provided are a best-effort approximation based on public resources \cite{ankrUsage}\cite{infuraUsage}.}
    \label{rpc-node-usage}    
\end{table}

\section{Problem Statement}\label{problem-statement}

There are no financial incentives for full non-validating, non-altruistic nodes in a permissionless network. An effective solution must address:

\begin{enumerate}
    \item \textbf{Incentives}: Implement on-chain crypto-economic incentives for non-transactional read RPC requests.
    \item \textbf{Scalability}: Scale to accommodate hundreds of billions of daily read requests due to the nature of read request volume.
    \item \textbf{Rate Limiting}: Facilitate rate limiting within a permissionless, multi-tenant environment.
\end{enumerate}

Current on-chain incentives are tailored for full validating or block building nodes, rewarding them for processing state transitions and collecting transaction fees. However, no analogous incentives exist for read-only operations, making reliance on altruistic node operators unsustainable for dApps.

\subsection{RPC Trilemma}

\begin{figure}
    \centering
    \includegraphics[width=0.7\textwidth]{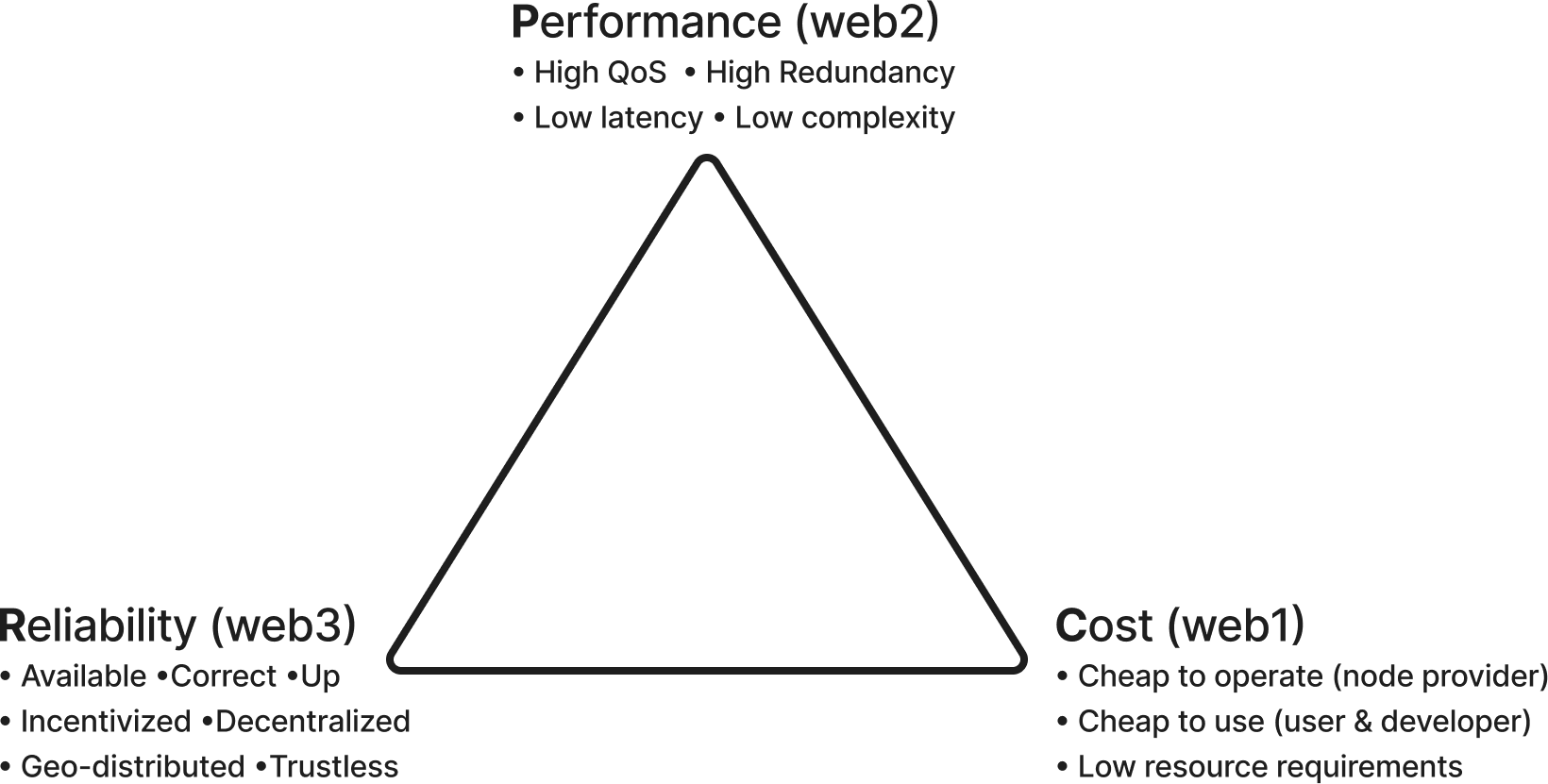}
    \caption{The RPC Trilemma} \label{fig:rpc-trilemma}
\end{figure}

This paper introduces the Relay Mining algorithm, addressing the aforementioned needs. With prospects for future expansion into tokenomics and the duality between centralized Gateways and permissionless hardware operators, we also outline the \textbf{RPC Trilemma}. Illustrated in Fig.~\ref{fig:rpc-trilemma}, this trilemma involves selecting RPC providers based on a trade-off among \textbf{Reliability}, \textbf{Performance}, and \textbf{Cost}. While our optimistic crypto-economic model incentivizes rate limiting and read request servicing, addressing just-in-time quality of service remains beyond this paper's scope.

\section{Experience \& Motivation}

The foundation of the problem statement in Section \ref{problem-statement} emerged from the challenges faced by a live permissionless network of RPC providers, notably Pocket Network. This network, built on Tendermint BFT \cite{tendermint}, began operation on October 22, 2021, and has served as a decentralized RPC provider, reaching a peak of 2 billion relays per day by May 8, 2023. Addressing the scalability of processing such a vast number of relays was the results of the following protocol characteristics:

\begin{enumerate}
    \item Utilization of a balanced (complete) Merkle Tree (IAVL) for relay book-keeping commitments. ~\cite{githubIAVL}
    \item Representation of each relay by a single leaf in the form of a signed (request, response) pair.
    \item Sorting of leaves in the Merkle Tree before insertion to ensure replay protection.
    \item Allocation of a significant portion (70\%) of each block to Merkle Proofs.
    \item Handling of hundreds to thousands of requests per second by each provider.
\end{enumerate}

These practices, especially the sorting of leaves for replay protection, introduced significant processing inefficiencies and security vulnerabilities \cite{alin2023sortedleaves}.

The Relay Mining algorithm proposes enhancements to address these inefficiencies within a permissionless and decentralized network of RPC service providers. Key improvements include:

\begin{enumerate}
    \item Implementation of a Sparse Merkle Sum Tree, which is more efficient in both storage and updates.
    \item Use of a hash collision mechanism to manage relay proofs, thereby decoupling tree size from demand growth.
    \item Employment of a ClosestMerkleProof proof-of-inclusion scheme to validate the work done without needing to sort the Merkle Tree.
\end{enumerate}

\section{Related Work}

\subsection{Rate Limiting Algorithms}

Most rate-limiting algorithms can be categorized as either Token Bucket algorithms (e.g., Leaky Bucket) or Window-based algorithms (e.g., Fixed Window, Sliding Window) \cite{betterprogrammingRateLimit}. Much of the existing work and literature that extends these to Distributed Rate Limiting (DRL) was captured in \cite{wustl2021drl} \cite{calu2007drl}, with two primary goals: 1. Finding a tradeoff between rate-limiting accuracy and communication overhead between brokers, 2. Reducing the DRL penalty on the end-user in exchange for accuracy. Although these approaches are valuable, they assume that the brokers or service providers, despite being geographically distributed, are not decentralized in terms of ownership and assume non-Byzantine guarantees.

\subsection{Distributed Rate Limiting}

Although the solutions discovered by the team at Washington University in St. Louis~\cite{wustl2021drl} do not directly apply to Relay mining, their formulation, as shown in Section \ref{rate-limiting-formulation}, is relevant. They distilled common Token Bucket rate-limiting algorithms into two key parameters when managing communication between \textit{Clients} and \textit{Brokers}:

\begin{itemize}
    \item[$\bullet$] \textit{r}: The rate at which tokens are generated. This bounds the long-term rate of message services.
    \item[$\bullet$] \textit{b}: The maximum number of tokens that can be accumulated. This bounds the maximum burst of messages that can be sent.
\end{itemize}

Their work on Scalable Real-Time Messaging (SRTM) encapsulated the following principles in reducing the Distributed Rate Limit (DRL) penalty:
\begin{itemize}
    \item[$\bullet$] \textit{Concentration}: Identification of the smallest number of brokers for message servicing to meet the network's Service Level Objectives (SLOs).
    \item[$\bullet$] \textit{Max-min}: Maximization of the minimum workload assigned to each broker.
    \item[$\bullet$] \textit{Correlation-awareness}: Assignment of publishers to brokers to minimize inter-publisher correlation and reduce message burstiness.
\end{itemize}

\subsection{Rate Limiting in Web3}

To the authors' knowledge, no other multi-tenant decentralized RPC node provider network currently operates in production. Centralized node operators such as Alchemy employ subscription tiers and compute units to manage rate limiting. This strategy, however, relies on centralization and does not necessitate on-chain proof and verification \cite{alchemyComputeUnits}. Compute units measure the total computational resources that applications utilize on platforms, rather than tallying individual requests. Another protocol in development, Lava, also intends to implement compute units for rate limiting \cite{lava2022}, but details of this implementation remain undisclosed. The lack of an existing decentralized RPC provider in production underscores the unique and complex challenges involved in maintaining decentralization while achieving on-chain proof and verification.

\section{Rate Limiting Formulation}\label{rate-limiting-formulation}

In order to present our approach to rate limiting in the context of Relay Mining, we will draw analogs from the concepts above to primitives in Pocket Network.

The primary actors of concern are:
\begin{itemize}
  \item[$\bullet$] \textit{Application}: The client sending RPC requests (real-time messages) to one or more Servicers (brokers).
  \item[$\bullet$] \textit{Servicers}: Distributed multi-tenant brokers responsible for handling and responding to RPC requests from the Application.
\end{itemize}

Parameters \textit{r} and \textit{b} map directly to on-chain values:
\begin{itemize}
  \item[$\bullet$] \textit{AppStake(r)}: A proxy into the maximum number of requests an Application can make across all current and future sessions across all valid Servicers. This directly relates to the Application's on-chain stake using the network's native token and can be increased through up-staking.
  \item[$\bullet$] \textit{RelaysPerSession(b)}: The maximum number of requests an Application can make across all Servicers in a single session.
\end{itemize}

The core SRTM principles map directly to on-chain parameters:
\begin{itemize}
  \item[$\bullet$] \textit{ServicersPerSession(Concentration}): The smallest number of Servicers that can handle an Application's requests in a specific session to meet the network's Service Level Objectives (SLO). These can include Quality of Service (QoS), decentralization, fairness, etc...
  \item[$\bullet$] \textit{RelaysPerServicerPerSession(Max-min)}: The maximum number relays a single Servicer can handle in a single session.
  \item[$\bullet$] \textit{SessionData(Correlation-awareness)}: Random assignment of an Application to a subset of valid Servicers for the duration of the session.
\end{itemize}

\section{Relay Mining}\label{relay-mining}

Relay Mining is a multi-step mechanism designed to ensure fair and permissionless distribution of work between a single Application and multiple Servicers. Utilizing data and parameters stored on the distributed ledger (i.e., on-chain data), this process provides incentive-driven rate limiting for the Application without necessitating additional communication between Servicers. A combination of commit-and-reveal mechanisms and probabilistically requested membership proofs is employed to verify the amount of work completed. Furthermore, a dynamic difficulty modulation is used to scale the network's capacity as it organically expands. Fig.~\ref{fig:session-lifecycle} illustrates the four stages of a Session Lifecycle.

\subsection{Session Lifecycle}

\begin{figure}
    \centering
    \includegraphics[width=\textwidth]{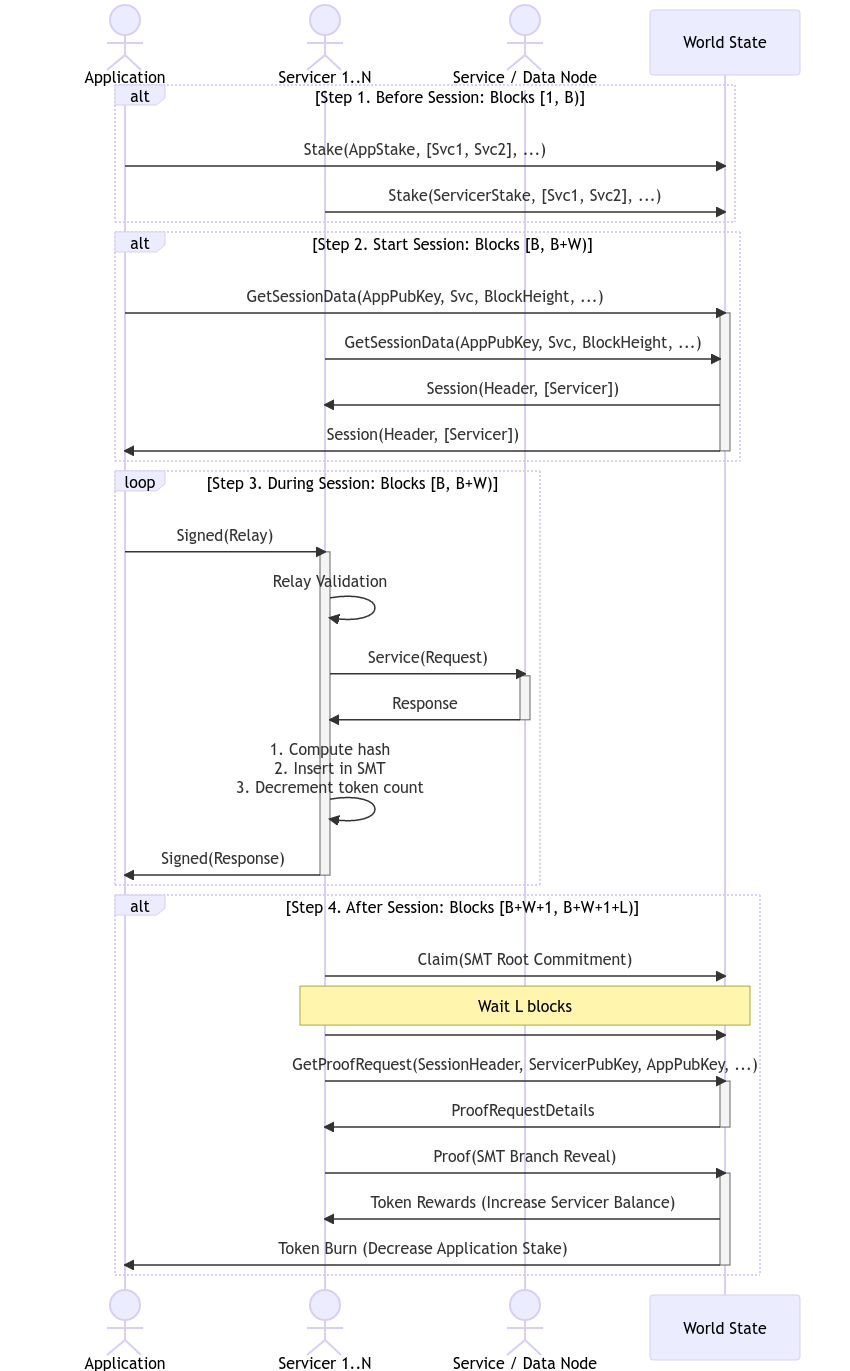}
    \caption{Interaction between the Application, Servicers and World State before, during and after a Session.} \label{fig:session-lifecycle}
\end{figure}

Applications and Servicers must stake a certain balance and the service(s) that they plan to access and provide, respectively. Note that the maximum long-term rate that the Application can receive is proportional to $r$. This is shown in Step 1 of Fig.~\ref{fig:session-lifecycle}.

Any Session (a periodic of time determined by a specific number of blocks) determines which Servicers can receive a reward for servicing an Application’s request. This is done with the assumption of the Random Oracle Model and uses on-chain data as entropy \cite{randomoracle1993} as shown in Step 2 of Fig.~\ref{fig:session-lifecycle}. Note that the maximum number of messages that can be serviced by each individual Servicer per session is $b$. Per the work described in \cite{wustl2021drl}, enforceable sub-buckets are not implemented and $r<\sum\limits_{l=1}^{k}r_l$. Due to the lack of guarantee that the sub-token count of each individual Servicer will be exhausted, a margin is provided, and a negative Application balance becomes theoretically possible. In this unlikely scenario, an Application would need to rotate the keys associated with its dApp in order to continue using the services provided by the network, and additional mitigations are reserved for future work.

\subsection{Payable Relay Accumulation}

Proposition 1 in \cite{wustl2021drl} demonstrated that in the context of highly dynamic workloads, splitting tokens into sub-buckets $(r_l, b_l)$ such that $r=\sum\limits_{l=1}^{k}r_l$ and $b=\sum\limits_{l=1}^{k}b_l$ can never improve the running sum of message delays. However, building on the work of \cite{calu2007drl}, who investigated probabilistic DRL, we can deduce that there is always an inherent tradeoff between absolute system accuracy and communication cost. Considering the possibility and added complexity of potential Byzantine actors in a multi-tenant, decentralized, and permissionless system, we opt for the tradeoff of accuracy in exchange for reducing coordination costs. In Algorithm~\ref{algo_relay}, $RelayAccuracy$ represents the maximum overhead that can cause an Application's stake to be negative at the end of a Session. This is shown in the first half of Step 3 in Fig.~\ref{fig:session-lifecycle}.

\begin{algorithm}[h]
    \caption{Payable Relay Accumulation}\label{algo_relay}
    \begin{algorithmic}[1]
        \State $AppStake \gets 10^6$\Comment{Amount of staked Applications tokens}
        \State $ServicersPerSession \gets 12$\Comment{Number of valid payable Servicers per session}
        \State $TTRM \gets 1000$\Comment{Token to relay multiplier per session}
        \State $RelayAccuracy \gets 0.2$\Comment{An accuracy error margin for TTRM}
        \State $b \gets \frac{AppStake \cdot TTRM}{ServicersPerSession}\cdot ( 1 + RelayAccuracy) $\Comment{Num token buckets per servicer per session}
        \newline
        \State Given a Session for a specific Application accessing Service Svc
        \newline
        \For{Servicers $[S_1, ..., S_N]$ between Blocks $[B, B+W)$}
            \State set $tokenCount = b$ for each Servicer
                \For{each relay R}
                    \State $signedRequest = Sign_{AppPrivKey}(SvcResponse)$
                    \If{$!valid(signedRequest)$}
                        \State continue
                    \EndIf
                    \If{$!payable(signedRelay, App, Servicer, Session)$}
                        \State continue
                    \EndIf
                    \If{$tokenCount \leq 0$}
                        \State continue
                    \EndIf
                    \State $SvcResponse = Svc(SvcRequest)$
                    \State $signedResponse = Sign_{ServicerPrivKey}(SvcResponse)$
                    \State $d = digest(signedRequest, signedResponse)$
                    \State $hasCollision = checkCollision(d, SvcDifficulty)$
                    \If{$hasCollision$}
                        \State $tokenCount -= 1$
                        \State $key = d$
                        \State $value = (signedRequest, signedResponse)$ 
                        \State $insertLeafInSparseMerkleTree(key, value)$
                    \EndIf
                \EndFor
            \EndFor
    \end{algorithmic}
\end{algorithm}

\subsection{Claim \& Proof Lifecycle}
The second part of Steps 3 \& 4 in Fig.~\ref{fig:session-lifecycle} provides an overview of the Claim \& Proof lifecycle implemented via a commit-and-reveal scheme.

\begin{figure}
    \centering
    \includegraphics[width=0.6\textwidth]{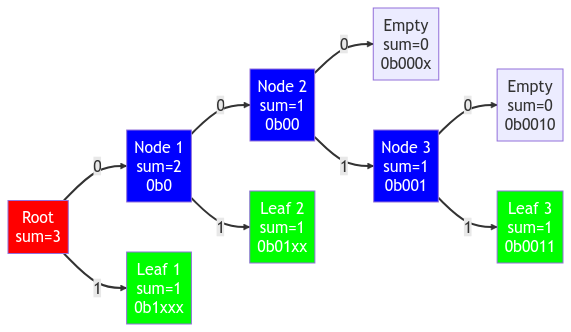}
    \caption{Four Bit Sparse Merkle Sum Trie} \label{fig:smt}
\end{figure}

Once a valid payable relay is serviced, assuming the $digest$ of $(SignedRequest, SignedResponse)$ using a collision-resistant hash function matches the difficulty for that chain/service as described in Section~\ref{volume-estimate}, it is committed for the purposes of volume estimation into a Merkle Tree Data Structure backed by a Key-Value storage engine. In order to avoid the compute overhead of tree re-balancing on every insertion, while maintaining efficient proof sizes, a Sparse Merkle Trie implementation \cite{diem2021jmt}, based on Diem's Jellyfish Merkle Tree sparse optimizations \cite{smt-github}, is used. The value stored on every leaf is the $Serialized(SignedRequest, SignedResponse)$ data structure, pointed to by the $digest$, whose bits act as the path to the leaf in the Trie. This approach guarantees space and compute efficiency, as well as preventing replay attacks through unique leaf guarantees without the need to maintain sorted leaves \cite{alin2023sortedleaves}. The only other modification that will be made to this Trie is that it will be implemented as a Sparse Merkle Sum Trie, so the sum of all the volume-applicable relays are committed to by the root \cite{plasma2019mst} as shown in Fig.~\ref{fig:smt}.

Several blocks after the Servicer submits a Claim to the tree Commitment via an on-chain transaction, there is a time window during which it must also reveal its proof of a randomly selected branch from the tree. Since the tree is not complete, the leaves are not indexed, and the set of all leaves is unknown to anyone but the Servicer's local state, a random branch must be requested through other means. Using on-chain data (e.g., Block Hash) determined after the commitment, a random set of bits equal to the key's length is generated in the Random Oracle Model \cite{randomoracle1993}, and the closest membership proof of an existing leaf that can be found must be revealed and submitted. For example, if the random set of bits is $0b0001$, the tree is traversed until it retrieves the membership proof for Leaf 3 in Fig.~\ref{fig:smt} with the key $0b0011$.

\subsection{Probabilistic Volume Estimation}\label{volume-estimate}

In a decentralized network where the exact count of processed relays is not tracked, we have to estimate the actual number of relays from sampled data. Consider the number of observed claims $C_b$ for a specific blockchain, denoted by $\mbox{bk}$, as a random variable following the binomial distribution:

\begin{equation}
    B(R_{bk}, p_{bk}) = \binom{R_{bk}}{C_{bk}} \cdot p_b^{C_{bk}} (1-p_{bk})^{R_{bk}-C_{bk}}, 
\end{equation}

Here, $R_{bk}$ represents the total number of processed relays and $p_{bk}$ denotes the hash collision probability given the current blockchain difficulty. We can then derive the value of $C_{bk}$ as follows~\footnote{For clarity, the subscript ${bk}$ is dropped in the remaining part of this work. Unless otherwise stated, all variables should be interpreted as \emph{per-blockchain}.}:

\begin{equation}
    R = \frac{C}{p}.
\end{equation}

This estimation is particularly accurate for large numbers of trials and higher probabilities~\cite{blumenthal1981estimating}, a scenario that is not difficult to achieve in the context of relay mining. The difficulty (and hence $p$) is controlled by the expected number of claims per block and the number of relays $R$ is typically very high ~\footnote{As per our observations, refer to section~\ref{network-stats} for details.}.

\subsection{Difficulty Modulation}\label{difficulty-modulation}

The proposed method for difficulty modulation follows Bitcoin's approach~\cite{nakamoto2008bitcoin}, which uses a proportional control of an average of the controlled variable to adapt the current hash collision probability~\cite{laanwj2023bitcoin}. 

The modulation of difficulty can be described by algorithm~\ref{algo_diff}. The method involves tracking the number of estimated total relays, denoted as $R_{ema}$, and modulating the hash collision probability $p$ to reach a target number of claims $T$. The \emph{Exponential Moving Average} (EMA) parameter $\alpha$ and the number of blocks between updates $U$ control the responsiveness of the algorithm, or the speed at which the difficulty adapts to new relay volumes.

The values assigned to these variables are not fixed and are estimated based on empirical observations. They are subject to change to accommodate the dynamic nature of relay volumes.

\begin{algorithm}[h]
    \caption{Relay Mining Difficulty Calculation}\label{algo_diff}
    \begin{algorithmic}[1]
        \State $T \gets 10^4$\Comment{Target claims by blockchain.}
        \State $\alpha \gets 0.1$\Comment{Exponential Moving Average Parameter.}
        \State $U \gets 4$\Comment{Number of blocks per difficulty update.}
        \State $R_{ema} \gets 0$ \Comment{Estimated blockchain relays, averaged by EMA.}
        \State $p \gets 1$ \Comment{Initial blockchain hash collision probability.}
        \State $\mbox{height} \gets 0$
    
        \While{True}
            \State $C \gets getAllClaims()$ \Comment{Get all relay claims.}
            \State $R \gets \frac{C}{p}$
            \State $R_{ema} \gets \alpha R + (1-\alpha) R_{ema}$
            \If{$\mbox{height} \% U$ == 0}
                \State $p \gets \frac{T}{R_{ema}}$ 
                \If{$p > 1$}
                    \State $p \gets 1$ \Comment{If total relays are lower than target, disable relay mining.}
                \EndIf
            \EndIf
            \State $\mbox{height} \gets +1$
        \EndWhile
    \end{algorithmic}
\end{algorithm}

\subsection{dApp Work Estimation}

A key component of the protocol method is the correct estimation of work performed by a dApp. This estimation is crucial for accurately processing the service payments from the dApp, which involves token burning, and for rewarding the Servicers, which involves token minting.\footnote{For simplicity, we assume a one-to-one mapping from an on-chain Application actor to a single dApp making RPC requests, though in theory, multiple dApps could leverage the key of a single Application.}

As explained in Section~\ref{difficulty-modulation}, the difficulty modulation relies on estimating the total relays per block on a given chain, where each chain can have multiple dApps. Given that the total number of relays $R$ is distributed among $A$ staked dApps, it is important to consider how edge cases could influence the bias and variability of the dApp relay volume estimation.

An experiment was conducted to examine this. Using a target of $T = 10^4$ claims per block, the variability and bias of the estimated number of relays $R_{dApp}$ were calculated. The difficulty $d$ (inverse of hash collision probability $p$) ranged from $1.25$ relays per claim to $1000$ relays per claim, and the dApp participation $v$ ranged from $0.1\%$ to $10\%$ of total blockchain traffic. For each test point, a total of $I=10^5$ draws from the resulting variable $x \sim B(R, v, p_b)$ were sampled. The bias and variability of the estimated dApp traffic were then calculated as follows:

\begin{equation}
    R_{\mbox{dApp}} = \sum_{i=1}^{I} \frac{x_i}{I},
\end{equation}
then:
\begin{equation}
    \label{equ:bias}
    \mbox{Bias} = \frac{R_{\mbox{dApp}} - (v\, R)}{(v\, R)} 100,
\end{equation}
and 
\begin{equation}
    \label{equ:var}
    \mbox{Variability} = 2 \, \sqrt\frac{\sum_{i=1}^{I}{(x_i-R_{\mbox{dApp}})^2}}{I} 100.
\end{equation}

\begin{figure}
    \centering
    \subfloat[\centering ]{{\includegraphics[width=0.46\textwidth]{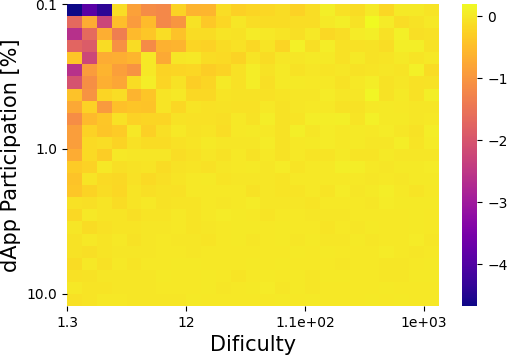} }\label{fig:dApp-bias}}%
    \qquad
    \subfloat[\centering ]{{\includegraphics[width=0.46\textwidth]{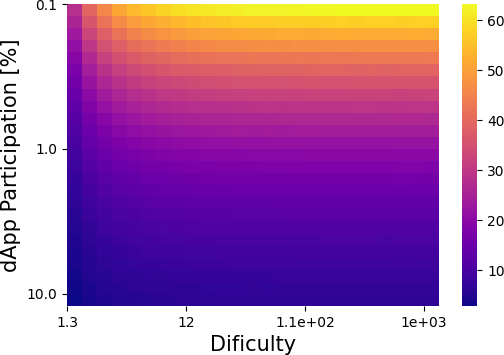} }\label{fig:dApp-var}}%
    \caption{Bias (left) and variability (right)  for each sample point in the experiment. The X and Y axis are in logarithmic scale.}%
\end{figure}

The resulting bias and variability are illustrated in heat maps in Fig.\ref{fig:dApp-bias} and Fig.\ref{fig:dApp-var}, respectively. We observe that the bias is close to zero across most of the sample space. Only when the number of $R_{\mbox{dApp}}$ is extremely low does bias become noticeable, but even then it remains above $-5\%$. Conversely, variability is higher, particularly when the dApp participation drops below $1\%$, under which conditions is grows rapidly.

\section{Experimentation on Pocket Network V0}

The proposed schema is expected to work under almost any kind of relay volumes and variations. This kind of relay volume variations can be easily obtained from the Pocket Network v0, as all relay data is tracked on its blockchain. Using this, it is possible to have an insight of how relay mining behaves on real traffic data and know the expected errors and bias of implementing it. 

The analysis include the bias, variation and transitory effects in the estimation of the blochchain traffic for four different served blockchains with different volume sizes and transitions. The used parameters were the same as presented in algorithm~\ref{algo_diff}. The main findings are summarized in table~\ref{table:v0tests}, the complete analysis is shown in the Appendix II. It can be seen that the transient effects on the target number of claims are significant but short lived, however the estimation of the total relay volume (responsible of controlling the difficulty) does not spike, ensuring an effective control of the proposed strategy.

\begin{table}
    \caption{Pocket Network V0 Experiment results. Response of the proposed method to different observed network changes.}\label{table:v0tests}
    \begin{tabularx}{\textwidth}{|C|C|C|C|C|C|}
    \hline
        &  Event Duration [blocks] & Min/Max Claims & Min/Max Target Error [\%] & Accumulated Target Error by Block [claims] & Min/Max Volume Estimation Error [\%] \\
    \hline
        Steady High Volume  & 3000 & 6789 / 20710 & -32.11 / 107.10 &   7 & -3.23 / 4.17 \\
        Soft surge          &  175 & 7804 / 14496 & -21.96 / 44.96  & 620 & -2.91 / 3.04 \\
        Step drop           &   44 & 1305 / 88963 & -86.95 / 789.63 & 170 & -1.92 / 3.45 \\
        Step surge          &   24 &  499 / 13897 & -95.01 / 38.97  & -60 & -5.21 / 2.59 \\
    \hline
\end{tabularx}
\end{table}

\section{Future Work \& Implementation}

Relay Mining sets the foundation for future advancements in the realm of decentralized rate limiting. Two potential avenues for further exploration include:

Compute Units: By assigning different weights to the leaves in the Merkle Sum Tree, rate limiting can be customized on a per-request basis, enabling more granular control over computational resources.

Websockets: Adapt Relay Mining to accommodate long-lived connections, such as websockets, by applying rate limiting to data transferred over the network at periodic snapshots rather than individual requests. This would enhance the versatility of the rate limiting approach to cater to a wider range of application requirements.

\section{Discussion \& Conclusions}

This paper introduces Relay Mining, a novel algorithm designed to scale efficiently to handle tens of billions of individual RPC requests daily. These requests are proven on a decentralized ledger via a commit-and-reveal scheme. The solution can account for the aggregate volume across all current Web3 Node RPC providers and comes equipped with an inbuilt difficulty modulation to dynamically adapt to the growing traffic in the Web3 industry.

Relay Mining builds upon the insights, performance, and data from the Pocket Network—a decentralized RPC network that has been operating a live mainnet for nearly two years at the time of writing. The network adeptly manages approximately 1.5 billion requests daily from over a dozen independent service providers servicing more than 50 blockchains. For secure and efficient insertions, and to meet random membership proof requirements, a Sparse Merkle Trees (SMTs) with probabilistic leaf insertions was adopted.

Relay Mining's design and mechanism incorporate principles from Scalable Real-Time Messages (SRTM) \cite{wustl2021drl}, aiming to alleviate Distribute Rate Limiting (DRL) penalties through concentration, max-min, and correlation awareness between Applications and Servicers. To our knowledge, this is the first DRL mechanism that leverages crypto-economic incentives to enable multi-tenant rate limiting capable of accounting for Byzantine actors.

This work demonstrates that the algorithm performs best under high relay volumes, a typical scenario in decentralized RPC networks. Although a slight bias exists for applications with exceptionally low comparative volume (less than $0.5\%$ of total blockchain traffic), this bias is minimal (under $|5\%|$). Furthermore, the estimation variance in these circumstances does not present a significant issue for the protocol itself, as it self-regulates with other low-traffic applications. Should the issue of extremely low-traffic applications become a concern, specific implementation mechanisms can be introduced to address it.

Finally, it is important to note that several relevant concepts in decentralized RPC networks—such as response integrity, Quality of Service enforcement, and extensions to general-purpose compute units—though outside the scope of this paper, are complementary to the problems that Relay Mining addresses.

\bibliographystyle{plain} 
\bibliography{bibliography}

\FloatBarrier


\newpage

\section*{Appendix I}

\subsection{Pocket Network Stats}\label{network-stats}

Since its genesis on October 22, 2021, up to the current date (May 10, 2023), Pocket Network has processed over 465 billion successful relays. During this period, the number of served relays per day increased from approximately 1.5 million to 1.5 billion, with a peak of 2 billion on May 8, 2023. The total relay traffic is composed of an increasing number of blockchains, currently supporting 50, distributed over 14 gateways, and over 12 major, with many minor, infrastructure providers across all continents. The performance from May 3, 2023, to May 10, 2023, on the dominant blockchains is presented in table~\ref{tab:stats-pokt}.

\begin{table}
    \caption{Main stats of the largest blockchains in the Pocket Network, from May 3rd, 2023 to May 10th, 2023. The data is averaged across all gateways, and distributed globally.}
    \label{tab:stats-pokt}
    \begin{tabularx}{\textwidth}{|C|C|C|C|C|C|C|}
        \hline Blockchain &  Share of Total Traffic [\%] & Volume [relays] & Avg. TPS & Peak TPS & Avg. Latency [ms] & Success Rate [\%] \\
        \hline
            Ethereum        &  $28.1$ & $365\times 10^6$ &  8300 &  16500 &  138 & 99.7039  \\
            Polygon Mainnet &  $25.3$ & $326\times 10^6$ &  3200 &   4700 &  130 & 99.7200 \\
            DFKchain Subnet &  $12.5$ & $164\times 10^6$ &  1800 &   2400 &  121 & 99.4618 \\
            Gnosis - xDai   &  $10.2$ & $132\times 10^6$ &  1500 &   2200 &  152 & 99.8337 \\
        \hline
    \end{tabularx}
\end{table}

The network is composed of 20,000 nodes; however, this number does not correlate with the number of deployed blockchain nodes. The current version of the protocol cannot verifiably prove the real number of blockchain nodes, as the Servicers are permissionless, but a rough estimate can be obtained through the nodes' service domains. The largest Servicer group has approximately 5,000 nodes (about 25\% of the network), followed by two others with around 2,500 nodes each (about 12.5\% of the network each). Then, the number of nodes by domain decreases. Around 90\% of the network is held by 14 independent domains. Assuming no blockchain node is shared among node runners~\footnote{This is a fair assumption given the competitive play between node runners.} and that each node runner has at least two nodes per staked blockchain~\footnote{Accounting for redundancy or geographical distribution of blockchain nodes.}, the Pocket Network has more than 30 independent blockchain servers on each of the main served blockchains. It is important to note that this is only an estimation; however, the number is significant in terms of decentralization and reliability.

\subsection{Network Actors}

\begin{figure}
    \centering
    \includegraphics[width=0.85\textwidth]{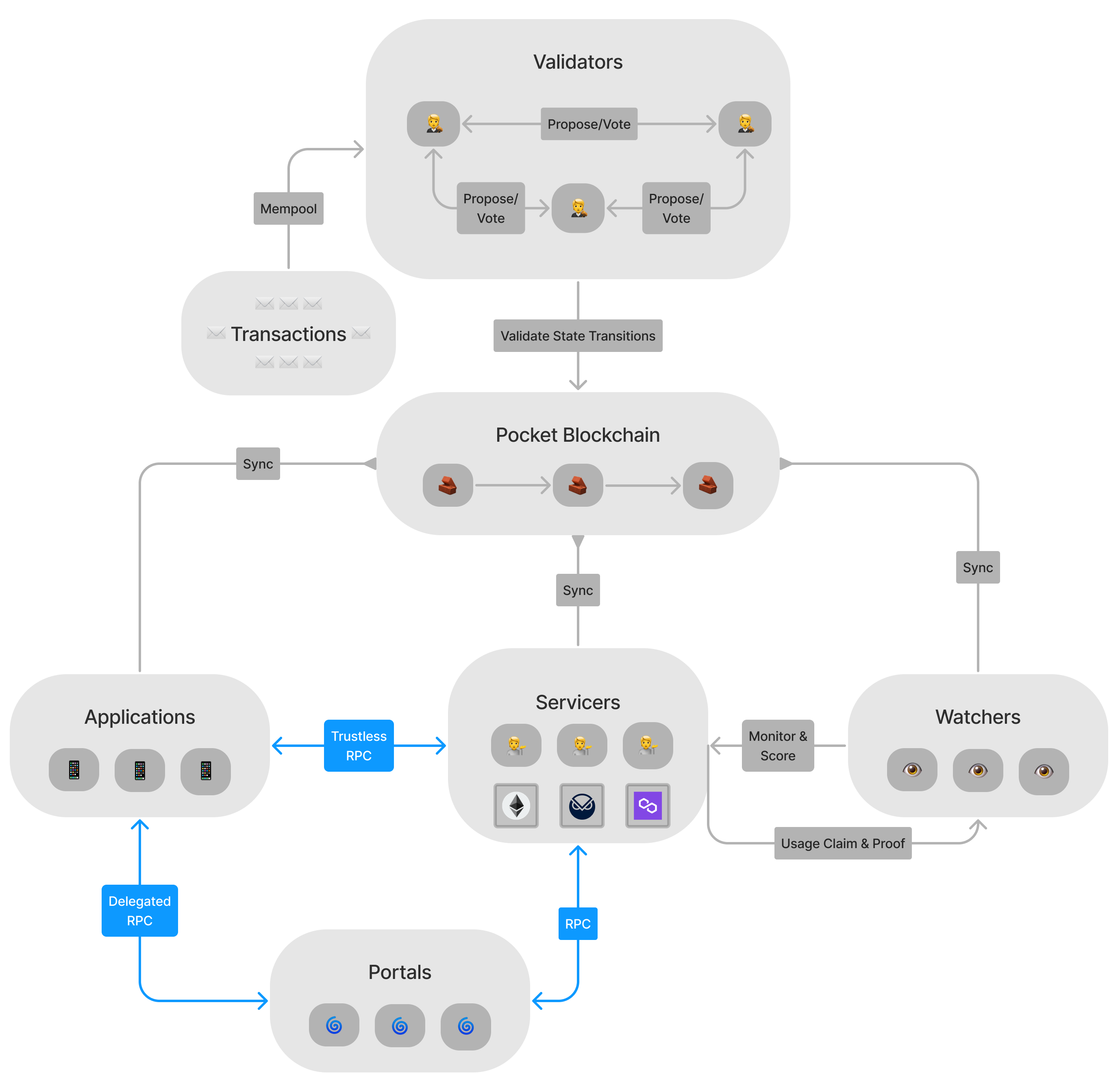}
    \caption{Pocket Network on-chain protocol actors in the next version of the protocol.} \label{fig:v1-actors}
\end{figure}

Fig.~\ref{fig:v1-actors} illustrates the interaction and responsibilities between the various protocol actors in Pocket Network. It is important to note that in the current, live iteration of the protocol, the \textit{Portals} and the \textit{Watchers} are off-chain and consolidated as a single entity. The roles of all the actors are outlined below. In the context of Relay Mining, the key item to note is that Applications pay for access to RPC services, and Servicers earn in exchange for handling those requests.

\begin{itemize}
    \item[$\bullet$] \textit{Validator} - Validates BFT state transitions of the distributed ledger
    \item[$\bullet$] \textit{Servicer} - Provides access to an RPC service and earns rewards in native cryptocurrency
    \item[$\bullet$] \textit{Watcher} - Makes periodic Quality of Service (QoS) checks of Servicers
    \item[$\bullet$] \textit{Application} - Pays for access to RPC service in native cryptocurrency
    \item[$\bullet$] \textit{Portal} - Optionally proxies requests on behalf of the Application and provides secondary off-chain services
\end{itemize}

\newpage
\section*{Appendix II}

\begin{figure}
    \includegraphics[width=\textwidth]{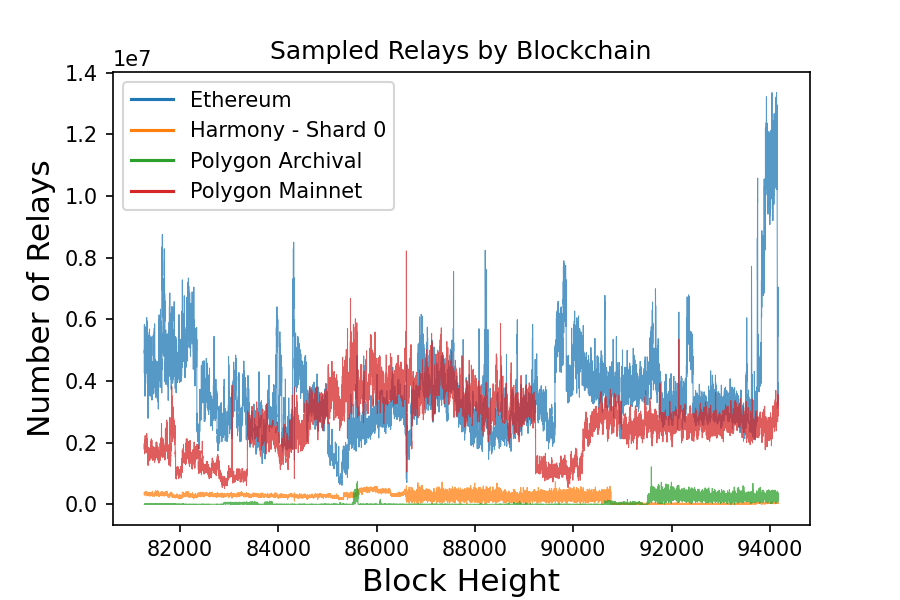}
    \caption{Evolution of the number of relays on the four selected networks, for a period of $\sim 134$ days.} \label{fig:v0-relays}
\end{figure}

The relay mining framework was simulated over real-world data obtained from the Pocket Network v0 blockchain. The simulation has the objective of estimating the deviations that are likely to be observed under different network behaviours, for this we selected for blockchains with different behaviours, under a period of $12898$ blocks ($\sim 134$ days). The selected networks were: Ethereum, Harmony Shard 0, Poligon Mainnet and Polygon Archival, the evolution of the traffic for each of them can be seen in Fig.~\ref{fig:v0-relays}. 

These cases were analyzed using the following configuration:
\begin{itemize}
    \item[$\bullet$] Target claims by block : $T = 10^4$
    \item[$\bullet$] Exponential Moving Average decay : $\alpha = 0.1$
    \item[$\bullet$] Number of blocks between updates : $U = 4$
\end{itemize}

\begin{figure}
    \includegraphics[width=0.9\textwidth]{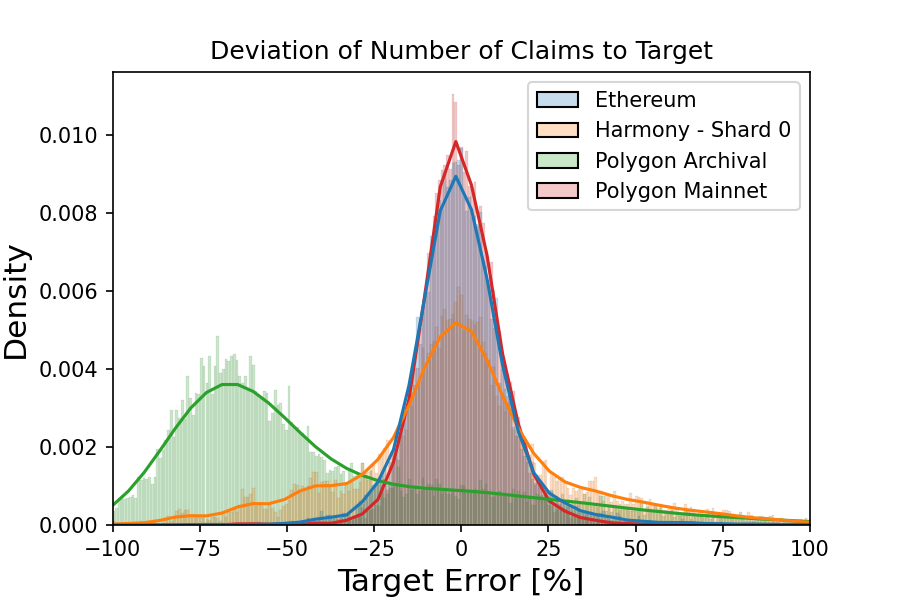}
    \caption{Distribution of the percentage deviation of the observed claims to the target number of claims ($100\, (C-T)/T$) on each of the observed blockchains, from block $81273$  to block $94171$. The Polygon Archival mean is lower than zero because the number of relays $R^*$ is below the target claims $T$ for a large part of the sample.} \label{fig:v0-target-error1}
\end{figure}

\begin{figure}
    \includegraphics[width=0.9\textwidth]{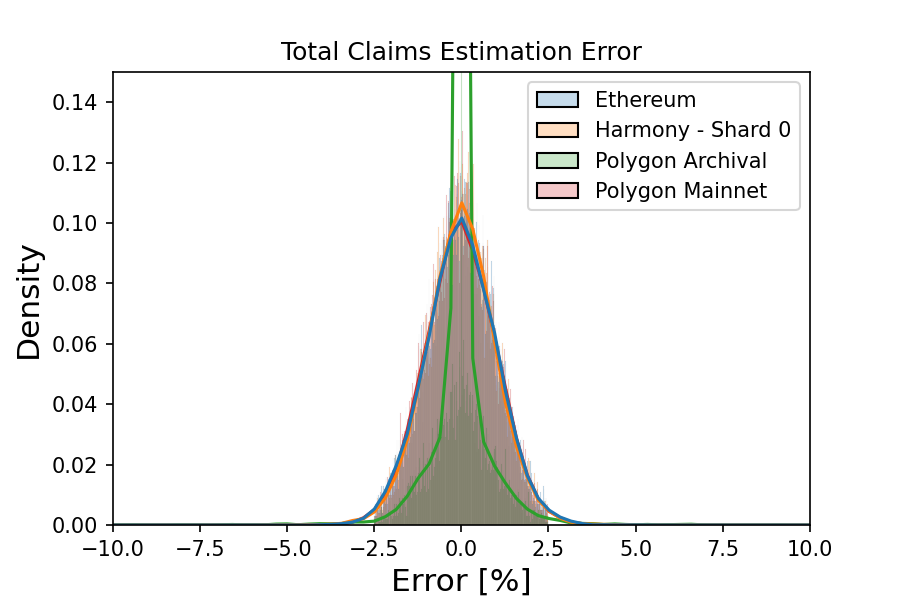}
    \caption{Distribution of the percentage deviation of the estimated relays to the real number of relays ($100\, (R-R^*)/R^*$) on each of the observed blockchains, from block $81273$  to block $94171$.} \label{fig:v0-target-error2}
\end{figure}

The process replicated the algorithm~\ref{algo_diff} to update the difficulty and calculated the errors in the number of claimed relays $C$ versus the selected target $T$, also the error of the real number of relays $R^*$ to the estimated number of relays $R$. Using this data it is possible to calculate the error to visualize the distribution of these errors in Fig.~\ref{fig:v0-target-error1} and Fig.~\ref{fig:v0-target-error2}, respectively.

It is also to analyze the transient effects on particular traffic behabiors. In the sampled blockchains data it is possible to observe four different cases:

\subsection{Steady High Volume - Polygon Mainnet}
\begin{figure}
    \includegraphics[width=0.9\textwidth]{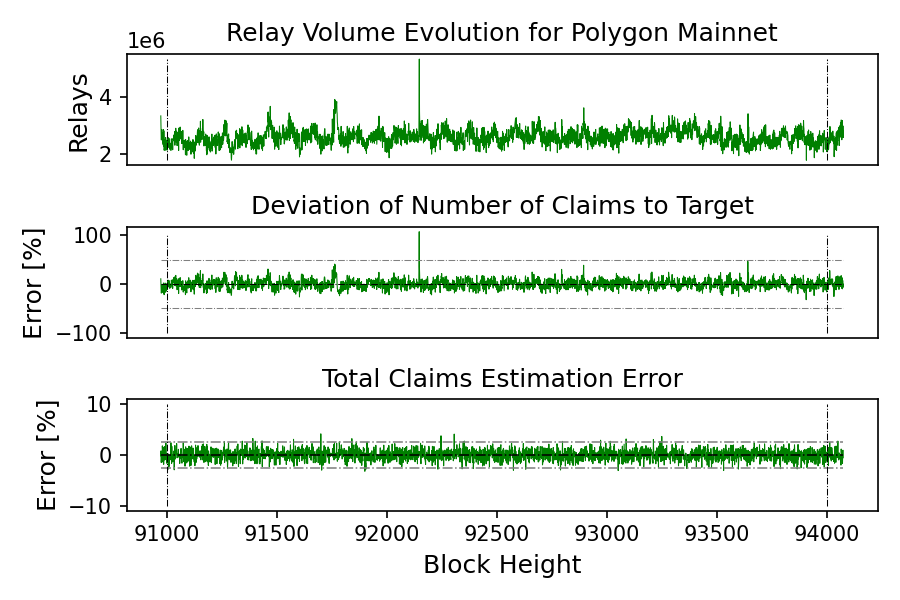}
    \caption{A sample of a steady high volume blockchain (Polygon Mainnet), from block $91000$  to block $94000$. From top to bottom, the evolution of the real number of relays, the percentage of deviation from target number of claims and the total relays estimation error. The horizontal dashed lines represent the zero error line (black) and the expected variation limits (grey). The vertical dashed lines limit the observed period.} \label{fig:v0-poly}
\end{figure}

This is the expected normal behaviour, where the number of relays by block stay almost constant or varies with low frequency. A blockchain that had this behaviour in the observed period of $3000$ blocks, from height $91000$ to $94000$ ($2023-04-08$ to $2023-05-07$) was the Polygon Mainnet. In Fig.~\ref{fig:v0-poly} a close-up of the process can be observed.

In this period the average target deviation was of $0.7\%$, with a peak target deviation of $|107 \%|$. The accumulated error resulted in a total of $20639$ extra claims, over $3000$ blocks. The average volume estimation error was $>0.00 \%$ with a peak of $|4.17| \%$.

\subsection{Soft Relay Surge - Ethereum}
\begin{figure}
\includegraphics[width=0.9\textwidth]{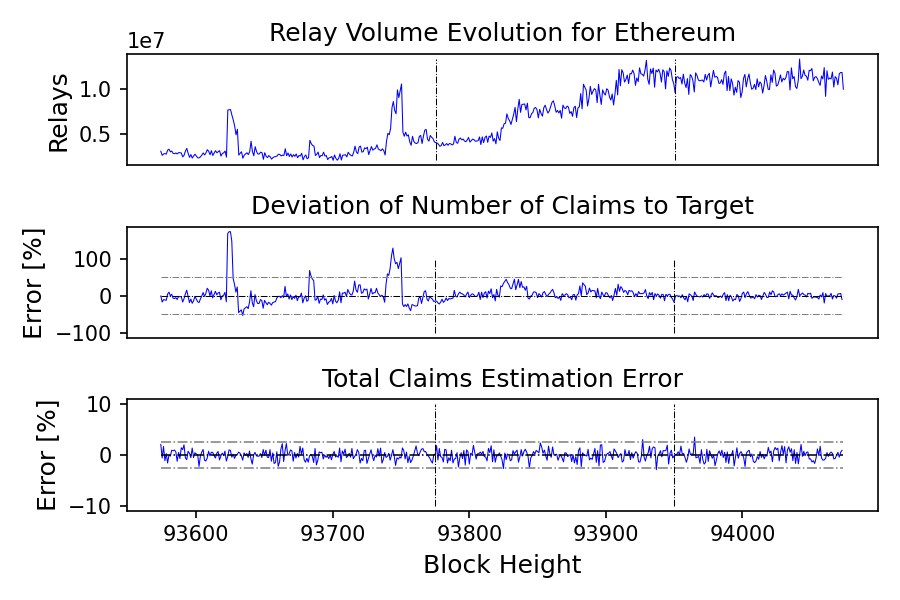}
\caption{A sample of a soft relay volume surge blockchain (Ethereum), from block $93775$  to block $93950$. From top to bottom, the evolution of the real number of relays, the percentage of deviation from target number of claims and the total relays estimation error. The horizontal dashed lines represent the zero error line (black) and the expected variation limits (grey). The vertical dashed lines limit the observed period.} \label{fig:v0-eth}
\end{figure}

To illustrate this case we use the \textbf{Ethereun} data over the blocks $93775$ to $93950$, this is from $2023-05-05$ to $2023-05-07$, during this period the volume of the blockchain processed in the Pocket Network raised from $2.9\times 10^6$ to $1.1\times 10^7$. The process was fast~\footnote{Compared to normal (non-step) increases observed in the network.} and steady. In Fig.~\ref{fig:v0-eth} a close-up of the process can be observed.

In this period the average target deviation was of $6.2\%$, with a peak target deviation of $|44.6 \%|$. The accumulated error resulted in a total of $108580$ extra claims, over $175$ blocks. The average volume estimation error was $-0.02 \%$ with a peak of $|3.04| \%$.

\subsection{Step Relay Drop - Harmony Shard 0}
\begin{figure}
    \includegraphics[width=0.9\textwidth]{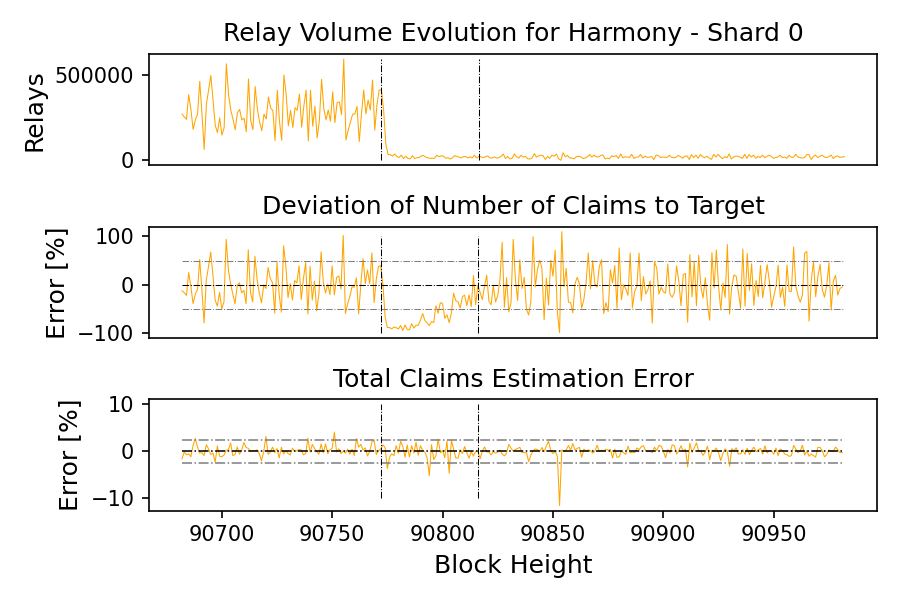}
    \caption{A sample of a step relay volume drop blockchain (Harmony Shard 0), from block $90772$  to block $90816$. From top to bottom, the evolution of the real number of relays, the percentage of deviation from target number of claims and the total relays estimation error. The horizontal dashed lines represent the zero error line (black) and the expected variation limits (grey). The vertical dashed lines limit the observed period.} \label{fig:v0-harmony}
\end{figure}

An step drop in relays occur when a major source of relays stop requesting data. This effect was observed at height $90772$ ($2023-04-05$) for the \textbf{Harmony Shard 0} blockchain, where the traffic dropped from $273\times 10^3$ to $21\times 10^3$ in less than four blocks. In Fig.~\ref{fig:v0-harmony} a close-up of the process can be observed.

In this period the average target deviation was of $60.2\%$, with a peak target deviation of $|95 \%|$. The accumulated error resulted in a total of $264717$ less claims, over $44$ blocks. The average volume estimation error was $-0.21 \%$ with a peak of $|5.21| \%$.

\subsection{Step Relay Surge - Polygon Archival}

\begin{figure}
    \includegraphics[width=0.9\textwidth]{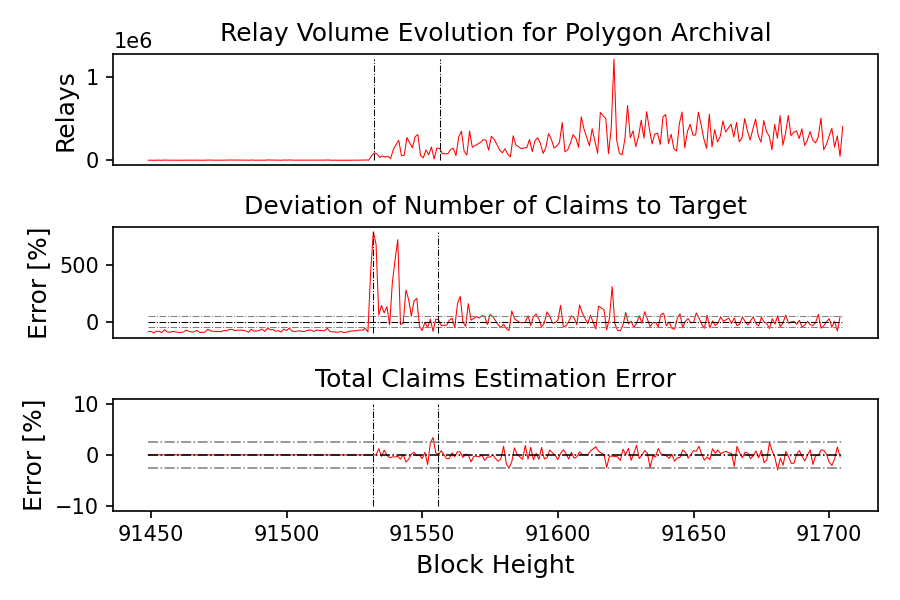}
    \caption{A sample of a step relay volume drop blockchain (Polygon Archival), from block $91532$  to block $91556$. From top to bottom, the evolution of the real number of relays, the percentage of deviation from target number of claims and the total relays estimation error. The horizontal dashed lines represent the zero error line (black) and the expected variation limits (grey). The vertical dashed lines limit the observed period.} \label{fig:v0-poly_arch}
\end{figure}

This case is typical when a new chain is introduced in the ecosystem, the evolution number of relays resembles an step function. This behaviour was observed recently, at height $91532$ ($2023-04-13$) for the blockchain \textbf{Polygon Archival}. This blockchain moved from $1120$ to $276\times 10^3$ relays per block. In Fig.~\ref{fig:v0-poly_arch} a close-up of the process can be observed.

In this period the average target deviation was of $170.32\%$, with a peak target deviation of $|789 \%|$. The accumulated error resulted in a total of $408767$ extra claims, over $24$ blocks. The average volume estimation error was $0.07 \%$ with a peak of $|3.45| \%$.

\end{document}